\documentclass[journal=ancac3,manuscript=article]{achemso}

\usepackage{morefloats}
\usepackage{color}
\usepackage{epsfig,graphicx,amsfonts,amsbsy}
\usepackage{amsmath,amsfonts,amsthm,amssymb}
\usepackage{appendix}
\usepackage{bbm}
\usepackage{makeidx}
\usepackage{url}
\usepackage{verbatim}
\usepackage{mathrsfs} 
\usepackage{morefloats}
\usepackage{appendix}
\usepackage{bbm}
\usepackage{makeidx}
\usepackage{url}
\usepackage{verbatim}
\usepackage{subcaption}

\usepackage[bookmarksnumbered,pdfpagelabels=true,plainpages=false,colorlinks=true,linkcolor=black,citecolor=black,urlcolor=black]{hyperref}

\usepackage{array}
\usepackage{booktabs}
\usepackage{multirow}
\usepackage{bbm}
\usepackage{tabularx}
\usepackage{cancel,soul}
\usepackage{nicefrac, xfrac}
\usepackage{ulem}
\usepackage{bbold}

\usepackage{xcolor}

\usepackage[version=3]{mhchem} % Formula subscripts using \ce{}
\newcommand{\bb}{\boldsymbol{{\rm b}}}
\newcommand{\bz}{\boldsymbol{{\rm z}}}

\newcommand{\bn}{\boldsymbol{{\rm n}}}
\newcommand{\ba}{\boldsymbol{{\rm a}}}
\newcommand{\bD}{\boldsymbol{{\rm D}}}
\newcommand{\be}{\boldsymbol{{\rm e}}}
\newcommand{\bS}{\boldsymbol{{\rm S}}}
\newcommand{\br}{\boldsymbol{{\rm r}}}

\newcommand{\bu}{\boldsymbol{{\rm u}}}

\def \fcfm {Departamento de F\'isica, FCFM, Universidad de Chile, Santiago, 8370448, Chile.}
\def \usach {Departamento de F\'isica, Universidad de Santiago de Chile, Santiago, 9170124, Chile.}
\def \cedenna {Centro  de Nanociencia y Nanotecnología CEDENNA, Avda. Ecuador 3493, Santiago, 9170124, Chile.}

\author{Martín Latorre}
\affiliation{\fcfm}
\email{martin.latorre@ug.uchile.cl}

\author{Joaquín Barra}
\affiliation{\fcfm}

\author{Juan Pablo Vera}
\affiliation{\fcfm}

\author{Joaquín Martinez}
\affiliation{\fcfm}

\author{Mario Castro}
\affiliation{\fcfm}

\author{Sebasti\'an Allende}
\affiliation{\usach}
\alsoaffiliation[\cedenna]{\cedenna}

\author{Alvaro S. Nunez}
\affiliation{\fcfm}

\title{Elastic Dislocation-based Skyrmion Traps: Fundamentals and Applications}


\begin{document}

\begin{abstract}
Topologically secure spin configurations, such as skyrmions and bimerons, offer a compelling alternative to conventional magnetic domains, potentially enabling high-density, low-power spintronic devices. These pseudo-particles, characterized by their swirling spin textures and nontrivial topological charges, are prevalent and notably influence their electronic, magnetic, and mechanical traits. This paper provides an in-depth overview of the interaction between a screw dislocation within a distorted magnetic lattice, exploring possible coupling mechanisms and establishing a promising link between two disparate topics in materials science: topological magnetism and topological elasticity. We first provide a classical analysis of skyrmion motion that reveals the dislocations as shallow traps on the magnetic texture. Afterwards, we provide an analysis of the quantized motion of the skyrmion and identify its quantum states.
We conclude by illustrating how the ideas in our paper can be implemented in simple yet compelling devices based on the shallow traps from an array of dislocations acting as frets in a race-track, controlling the motion with a low current activation mechanism.
\end{abstract}



%\tableofcontents



\section{Introduction}
\label{introduction}

Spintronics\cite{Dey2021, Guo2022}, an expanding research area, continues to explore innovative methods for controlling electron spin to enhance data storage, and processing. Topologically protected spin configurations\cite{Fert2017, Li2023, Tokura2020, Ji2024, Fullerton2025, Cacilhas2018}, such as skyrmions and bimerons, are promising candidates due to their stability, nanoscale dimensions, and efficient motion under the influence of current. These quasi-particles, characterized by their swirling spin configurations and non-trivial topological charges, offer a compelling alternative to conventional magnetic domains, potentially enabling high-density, low-power spintronic devices\cite{Koshibae2015, Luo2021, Sisodia2022, Zhao2024}. Skyrmions, generally occurring in systems with perpendicular magnetization, exhibit cylindrical symmetry and a topological charge of $q=\pm 1$\cite{Nakahara2003, Zang2018}.  The ability to generate, adjust, and systematically alter these unique topological spin textures is crucial for their use in spintronic systems.

Simultaneously, the mechanical properties of materials, particularly the presence and nature of crystalline defects such as dislocations\cite{Anderson2017, Caillard2007, Friedel2013}, are crucial for device performance and reliability. Dislocations, line defects in the crystal lattice of materials, are prevalent and significantly influence their electronic, magnetic, and mechanical properties\cite{Chen2017, Han2025}. For example, they may function as electron-scattering centers, influencing electrical conductivity, or as pinning sites for magnetic domain walls, affecting their motion\cite{Kaappa2024, Kurtzig1970}. In spintronics, understanding the impact of defects is crucial as device scales decrease, underscoring the significant interplay between structural integrity and spin phenomena. Recent research has shown that dislocations can trigger spin-orbit coupling\cite{Hu2018} or generate localized magnetic moments\cite{Carpio2008}, indicating a more intricate and potentially advantageous association than previously considered\cite{Saji2025, Yamada2022}.

The independent exploration of topological spin textures, such as skyrmions and spin waves\cite{Turski2009, Gestrin2012, Azhar2022}, alongside crystalline defects, such as dislocations, is well recognized. However, the direct interplay between these core elements of condensed matter physics is largely under-investigated. This gap offers a compelling scientific challenge and is crucial for advancing the design and optimization of next-generation spintronic devices. What is the interaction between mobile dislocations traveling through a material and the stable, localized spin configurations of skyrmions? Could dislocations act as either conduits or obstacles for the movement of these topological defects? Conversely, might skyrmions affect dislocation mobility or formation, potentially enabling novel strategies in strain engineering or defect management?

Magnetic skyrmions, originally conceived as classical topological spin textures, also exhibit intriguing quantum properties that emerge at the nanoscale\cite{Petrovi2025}. In the quantum regime, the collective spin configuration of a skyrmion behaves as a quantized quasiparticle, possessing discrete energy levels associated with its translational and internal degrees of freedom. Quantum fluctuations can lead to tunneling between topological sectors, enabling phenomena such as skyrmion creation and annihilation through quantum nucleation processes\cite{RoldanMolina2015, Ornelas2025}. Moreover, the Berry phase associated with the skyrmion’s spin texture gives rise to emergent electrodynamics, where skyrmions acquire an effective magnetic flux\cite{RoldanMolina2016, Zou2025} and obey nontrivial quantum statistics. These effects open the possibility of treating skyrmions as quantum bits or carriers of spin information in future quantum technologies, where their topological protection could be harnessed to achieve robust quantum control and coherence.

This paper offers an in-depth overview of the interaction between dislocations and skyrmions. We examine the core principles governing dislocations and topological spin textures, and explore possible interaction mechanisms. Key aspects include the stress fields from dislocations, their impact on magnetic anisotropy, the Dzyaloshinskii-Moriya interaction (DMI), which is essential for the stability of skyrmions, and potential spin-lattice coupling. By integrating existing insights and identifying crucial questions, this work aims to support future theoretical and experimental studies of their interactions and implications for spintronic technologies. Understanding these interactions fully is critical in advancing topological spintronics.
\section{Characterization of the dislocation's vicinity}
The interplay between topological magnetic configurations, such as skyrmions, and crystal dislocations is a critical area of investigation in spintronics and condensed matter physics, attracting significant attention in recent research\cite{Saji2025, Turski2009}. The central aim of this examination is to unravel the influence of crystal lattice defects and the stable arrangements of magnetic moments on a material's physical properties. Additionally, this study enables the detection of phenomena arising from these interactions. To accurately characterize these configurations, it is essential to utilize magnetoelastic couplings. These couplings facilitate the identification of the impact of a dislocation-induced deformation of a material's crystal lattice on the crystal's magnetic characteristics, and vice versa, thereby providing a deeper understanding of the mutual influences.\\
Accordingly, we consider a screw dislocation and its interaction with a skyrmion tube. The dislocation is taken along  $\hat{z} $ characterized by a Burgers vector $\bb =  b \bz$\cite{Landau1984, Kleinert2007, kleinert1989}. Outside the core, the elastic displacement is $u_z = b/(2\pi) \theta$ in cylindrical coordinates, producing long-range torsional shear and an azimuthal shear strain that decays as $1/\rho$: $
\sigma_{rz}=\sigma_{zr}=\frac{\mu b}{2 \pi \rho}
$.
These inhomogeneous fields act as slowly varying, defect-centered perturbations that can bias the width, orientation, and chirality of skyrmions over distances well beyond the lattice spacing.  To describe the coupling of the elasticity and micromagnetic theory, we use the vielbein (triad) $B_{i}^{\alpha}$\cite{Bausch1999, Zee2013, Wald1984}, which maps the crystal direction $\be_{\alpha}$ to lab coordinates $x^{i}$ and induces the effective metric $g_{ij}= B_{i}^{\alpha}  B_{j}^{\beta}\delta_{\alpha\beta}$ (see Appendix A for details). 

Under this mapping, derivatives are taken with respect to the deformed geometry\cite{Zee2013, Wald1984}. For a scalar field, the covariant derivative reduces to the partial derivative, $\nabla_i \phi = \partial_i \phi $. In contrast, for a vector field, the covariant derivative reads $\nabla_iV^{\nu} = \partial_i V^{\nu} +  \Gamma_{kj}^{\nu} V^{j}$, where $\Gamma_{kj}^{\nu}$ is the affine connection. 

\section{Theory of the magnetization state next to a dislocation}
In our model, the skyrmion tube is stabilized solely by exchange and Dzyaloshinskii–Moriya interactions (DMI). In this context, we focus on non-centrosymmetric bulk chiral magnets where the bulk DMI (with energy density of the B20 form) stabilizes Bloch-type textures, chiefly the B20 family (MnSi, FeGe, MnGe) and the insulating Cu$_2$OSeO$_3$. In these systems, first-neighbor exchange  $J$ typically lies in the few tens of meV, while the Dzyaloshinskii–Moriya constant $D$ is usually smaller with magnitude $D$ $\approx 10^{-2}-10^{-1}$ meV \cite{Lin2014, Camley2023, Borisov2021}, yielding $D/J \sim 10^{-3}-10^{-1}$. Additional contributions to the energy, such as the Zeeman or anisotropy energies, contribute to the overall shape of the skyrmion but do not couple to its positional degrees of freedom.  
To study the dynamics of the skyrmion tube under elastic deformation, we will derive an effective Thiele equation using a Lagrangian formalism. The Lagrangian takes the form $L = T - E$, where $T $ contains the dynamical contributions: the Berry phase term and the adiabatic spin-transfer torque arising from a spin-polarized electric current \cite{Bazaliy1998, FernndezRossier2004,Li2003}: $T = \dfrac{\hbar S}{a^3}\int d\br A(\bS)\cdot[\partial_t \bS + (\bu\cdot \nabla )\bS]  $ with $\bu = \dfrac{P a^3 \mathbf{J}_e}{2 e S}$, where $P$ is the spin polarization, $e$ is the electron charge, and $\mathbf{J}_e$ is the applied current density. The energetic contributions are encoded in $E$, which includes the contributions of the exchange and DMI. Dissipative effects, such as Gilbert damping and the non-adiabatic spin-transfer torque, are incorporated through a Rayleigh dissipation functional given by $\mathcal{R} = \dfrac{\hbar  S}{2 a^3} \alpha \int d\br [\partial_t \bS + \dfrac{\beta}{\alpha} (\bu\cdot \nabla \bS)]^2$, where $\alpha$ is the damping constant and $\beta$ is a parameter that quantifies the degree of
non-adiabaticity \cite{Tatara2008}. 

We begin by considering the discrete Heisenberg exchange on the distorted lattice:
\begin{align}
    H_{EX}=\frac{JS^2}{2}\sum_{\bn,\ba(\bn)}\bS(\bn)\cdot \bS(\bn+\ba(\bn)) 
\end{align}
where the sum is taken over the lattice sites $\bn$ and the nearest-neighbour vectors $\ba(\bn)$.  Following the procedure used by R. Bausch et al. \cite{Bausch1998,Bausch1999} for deriving continuum limits in distorted magnetic lattices, we expand the discrete Heisenberg exchange to leading order in a gradient expansion. In the specific case of a screw dislocation, the torsion tensor satisfies $\nabla_k T_{ij}^{k} = 0$. As a consequence, the covariant derivative acting on the spin field reduces to the ordinary partial derivative, and the exchange energy assumes the covariant continuum form:
\begin{equation}
    H_{EX}=-\frac{JS^2}{2}\int d^2x\sqrt{g}g^{ij}\nabla_i\bS\cdot \nabla_j\bS,
\end{equation}
where $g = \det(g_{ij})$ and $g^{ij} = (g_{ij})^{-1}$ is the inverse of the metric tensor. It is important to note that this expression compactly encodes the influence of local lattice deformations induced by the screw dislocation, through the metric tensor $g_{ij}$. For the Dzyaloshinskii–Moriya interaction, we start from the lattice form \cite{Niu2024, Robertson2020}:
\begin{align}
H_{DMI}=S^2\sum_{\bn,\ba(\bn)}\bD\cdot[\bS(\bn)\times\bS(\bn+\ba(\bn))]
\end{align}
where $\bD$ is the Dzyaloshinskii–Moriya vector which is consider $\bD = D \hat{\ba}(\bn)$ for bulk DMI and $\bD = D (\bz\times \hat{\ba}(\bn))$ for the interfacial case. By applying the same procedure as those used for the exchange interaction, this term can be expressed in the continuum limit as:
\begin{equation}
    H^{bulk}_{DMI}=\frac{DS^2}{a}\int d^2x\sqrt{g}g^{ij}\be_i\cdot(\bS\times \nabla_j \bS)
\end{equation}
where $\be_i=B_i^\alpha \be_\alpha$ represents the directional vielbein of the system. The analysis for interfacial DMI is presented in Appendix B.

In what follows, we choose $\varepsilon=J S^2$ as the unit of energy, $\tau=\hbar/DS$ as the unit of time, and $\ell=a\,J/D$ as the unit of length. 
To describe the magnetization texture, we employ a stereographic projection method\cite{Garg2003, Timofeev2022, Horley2009}, which maps the unit sphere from its north pole to the complex plane. This construction enables a compact and analytical representation of the magnetization vector field, particularly useful for topological solitons. Under this mapping, the spin components are expressed as:
\begin{equation}
    \begin{aligned}
        S_x &= \frac{\psi+\bar{\psi}}{1+\psi \bar{\psi}} & \quad
        S_y &= i\frac{\bar{\psi}-\psi}{1+\psi \bar{\psi}}& \quad
        S_z &= \frac{1-\psi \bar{\psi}}{1+\psi \bar{\psi}},
    \end{aligned}
\end{equation}
where $\psi=\psi(\zeta,\bar{\zeta},t)$ is a complex scalar field and $\zeta =x+iy$,  $\bar{\zeta} =x-iy$ are the complexified planar coordinates. Under this approach, the skyrmion's structure is encoded in the scalar field $\psi$. 

\begin{figure}[H]
    \centering
    \begin{subfigure}[b]{0.49\linewidth}
        \centering
        \subcaption*{(a)} % <--- solo letra arriba
        \includegraphics[width=\linewidth]{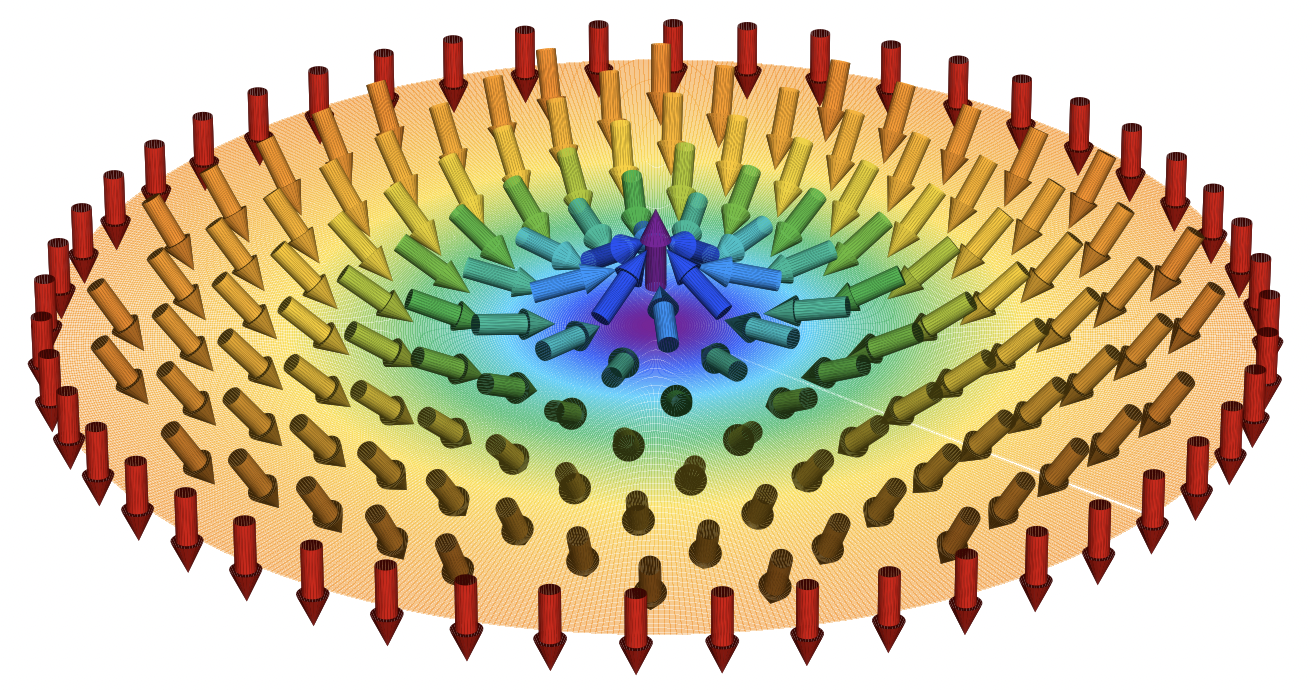}
        \label{fig:no_disi}
    \end{subfigure}
    \hfill
    \begin{subfigure}[b]{0.43\linewidth}
        \centering
        \subcaption*{(b)}
        \includegraphics[width=\linewidth]{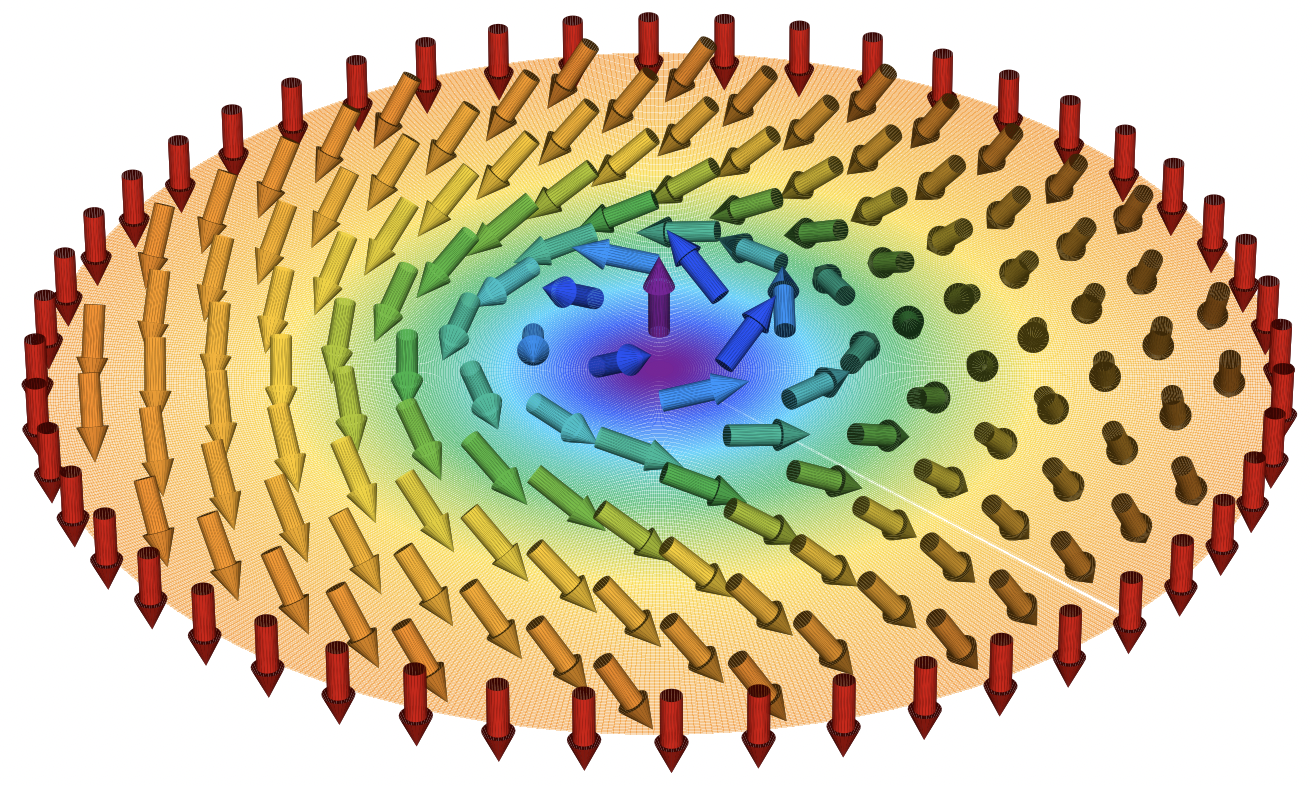}
        \label{fig:disi}
    \end{subfigure}
    \caption{(a) shows a Néel-type skyrmion, in which $A$ is a real number. (b) corresponds to a Bloch-type skyrmion, where $A$ is a pure imaginary number.}
    \label{fig:comparacion4}
\end{figure}

A minimal ansatz capturing a single skyrmion configuration is given by
$\psi(\zeta,\bar{\zeta},t) = A(t)(\zeta - w(t))$,
where $w(t)$ denotes the skyrmion center and $A(t)=|A(t)|{\rm e}^{i\gamma(t)}$ is a complex parameter that combines the core size $|A|^{-1}$ and the helicity $\gamma$, which characterizes the internal rotation of the texture. This form ensures that $S_z(\psi\to \infty) = -1$ and $S_{z}(\psi = 0) = 1$, describing adequately both Neel and Bloch skyrmion as in shown in figure \ref{fig:comparacion4}.(a) and \ref{fig:comparacion4}.(b), respectively. This framework not only simplifies the representation of topologically non-trivial textures but also enables analytical computation of micromagnetic energy contributions under deformations. Furthermore, it reveals natural collective coordinates, such as the skyrmion position
 z(t), core size $|A(t)|^{-1}$, and helicity $\gamma(t)$ that can be treated as dynamical variables within effective field theories.

%To study the energy response to the dislocation, we first express the inverse metric in complex coordinates, as seen in Appendix A.
\section{Interaction Between Skyrmions and Dislocations}
To analyze the energetic stability and deformation response of a skyrmion tube in the presence of a screw dislocation, we insert the ansatz into the continuum energy functionals derived above. Substituting the projected magnetization components into the covariant exchange term yields:
\begin{equation}
    H_{EX}(\zeta,\bar{\zeta})=\int i
    d\zeta\wedge d\bar{\zeta} \frac{\sqrt{g}g^{\bar{\zeta}\zeta}\partial_{\bar{\zeta}}\bar{\psi}\partial_{\zeta}\psi}{(1+\psi\bar{\psi})^2}
\end{equation}
Here $g^{\bar{\zeta}\zeta}$ is the component of the inverse metric in complex coordinates, and can be related to the cartesian components through $g^{\bar{\zeta}\zeta}(\zeta,\bar{\zeta})=g^{xx}(\zeta,\bar{\zeta})+ig^{xy}(\zeta,\bar{\zeta})-ig^{yx}(\zeta,\bar{\zeta})+g^{yy}(\zeta,\bar{\zeta})=2$. 
In the same way, it is possible to apply the stereographic variable transformation to the bulk and the interstitial DMI exchange, obtaining:
\begin{align}
 %   H_{DMI}^{bulk}=\frac{D}{2a}\int id\zeta\wedge d\bar{\zeta}\left[ \frac{big^1 \partial_\zeta\psi}{2\pi \zeta(1+\psi\bar{\psi})^2}+\frac{2i\partial_\zeta\psi}{(1+\psi\bar{\psi})^2}+\frac{2g^1\bar{\psi}\partial_\zeta\psi}{(1+\psi\bar{\psi})^2}+\frac{big^1 \bar{\psi}^2\partial_\zeta\psi}{2\pi\bar{\zeta}(1+\psi\bar{\psi})^2}+h.c.\right] && \\ 
        H_{DMI}^{bulk}=\int id\zeta\wedge d\bar{\zeta}\left[ \frac{2i\partial_\zeta\psi}{(1+\psi\bar{\psi})^2}+g^1\frac{\left(2\bar{\zeta}\zeta\bar{\psi}+ib(\bar{\zeta}+\zeta\bar{\psi}^2)\right) \partial_\zeta\psi}{2\pi \bar{\zeta}\zeta(1+\psi\bar{\psi})^2}+h.c.\right]
\end{align}
where $g^1$ is a linear combination of the components of the inverse metric in complex coordinates; that is: 
$
g^1(\zeta,\bar{\zeta})=g^{zy}(\zeta,\bar{\zeta}) - i g^{zx}(\zeta,\bar{\zeta})=-\frac{b}{2\pi\bar{\zeta}}$.

\begin{figure}[H]
    \centering
    \includegraphics[width=0.99\linewidth]{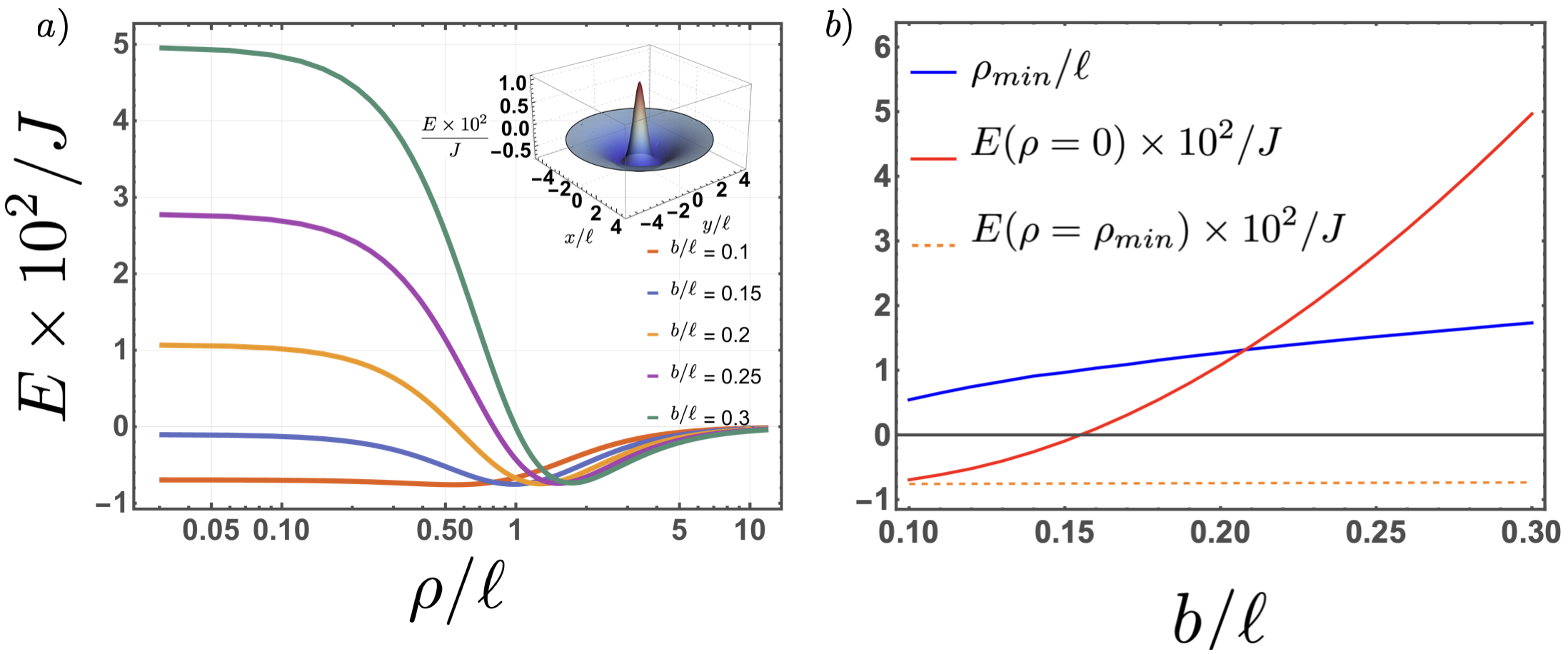}
   \caption{a) Normalized energy $E/J$ as a function of the skyrmion position $\rho/\ell$, considering $A=i$, $\gamma=\pi/2$, and $J/D=10$. The inset shows the normalized energy $E/J$ for $b/\ell=0.2$ as a function of the skyrmion coordinates $x/\ell$ and $y/\ell$, showing the effect of the phase. b) Normalized energy for $\rho=0$ and $\rho = \rho_{min}$ as a function  $b/\ell$. The blue line shows the dependence of the equilibrium radius $\rho_{min}$ on $b/\ell$. }
    \label{fig:combined}
\end{figure}

To analyze the influence of the dislocation on the skyrmion energy landscape, we evaluate the total energy of the system defined as $E/J=E_{EX}+E^{bulk}_{DMI}$. Notably, the exchange energy remains invariant under the presence of a screw dislocation, yielding $E_{EX}=4\pi$, since the skyrmion exchange energy is unaffected by the defect. In contrast, the Dzyaloshinskii–Moriya   contribution becomes sensitive to the distortion encoded in the geometry and cannot be evaluated analytically in closed form.  To examine the resulting interaction, we numerically compute the normalized energy. The results, depicted in Fig. \ref{fig:combined}, reveal how the total energy varies as a function of the normalized skyrmion position $\rho/\ell$ for different values of the normalized Burgers vector $b/\ell$. The inset illustrates the energy landscape for $b/\ell = 0.2$, revealing a radially symmetric confinement profile centered at the defect core. The potential resembles a ``Mexican hat'' shape, with a minimum at a finite distance from the core and a steep energy rise near the dislocation line. This indicates that the dislocation can acts as an isotropic pinning center and confines the skyrmion in a potential well with no preferred angular direction.

%This enables the characterization of the interaction between skyrmions and dislocations.
%To study the impact of the dislocation on the skyrmion, it is essential to examine how the interaction modifies the system’s energy. For this purpose, we consider that the total energy is $E/J=E_{EX}+E^{bulk}_{DMI}$.\\
%It is worth noting that $E_{EX}=4\pi$ for a screw dislocation, since the skyrmion exchange energy is unaffected by the defect. In contrast, the expression for the Dzyaloshinskii–Moriya (DM) exchange cannot be solved analytically. Therefore, the normalized energy term $E/J$ was plotted, allowing the observation of variations in the total energy:

\section{Classical theory of dislocation-based skyrmion traps}
\begin{figure}[H]
    \centering
    \includegraphics[width=1.02\linewidth]{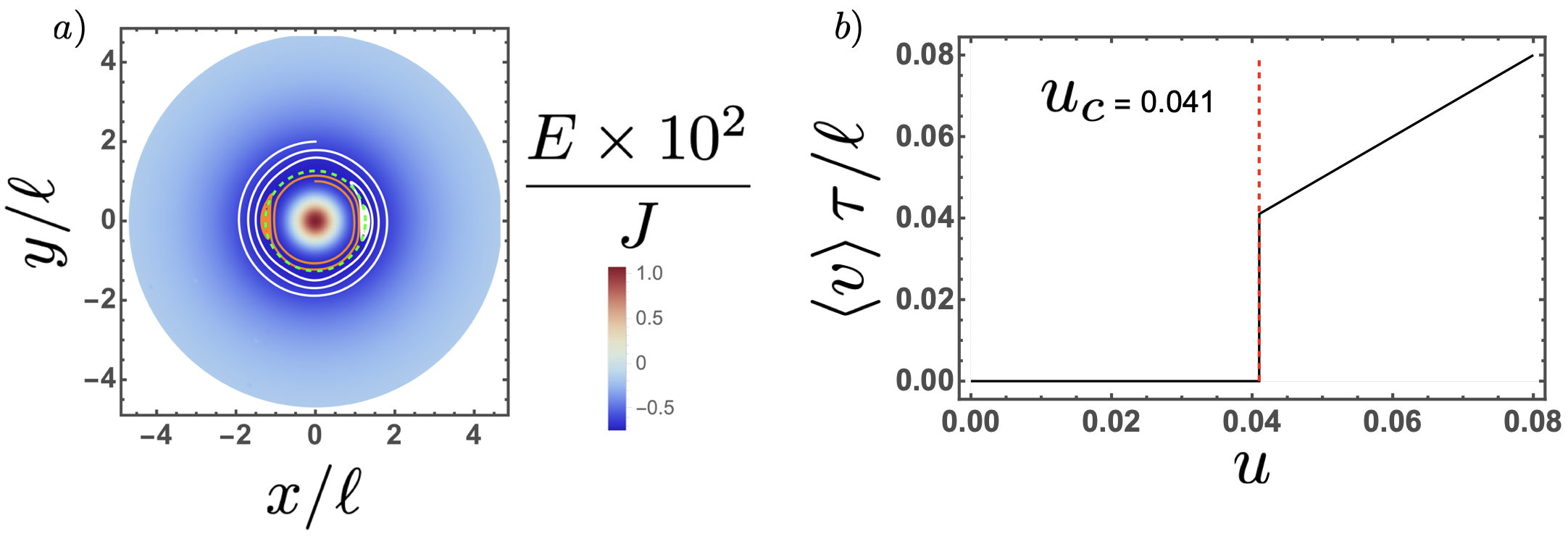}
    \caption{a) Trajectory of the skyrmion with an initial position larger (white) and smaller (orange) than the equilibrium radii $\rho_{min}$ (dashed green circle). Considering $b/\ell=0.2$, $A=i$, $\gamma=\pi/2$, $\alpha= \beta =0.02$, and $J/D=10$.  The surface represents the energy $E$ in the 2D plane. b) Average long-term velocity as a function of the current $u$. A trapping mechanism is revealed where the terminal velocity vanishes for currents up to a critical value. Restoring the S.I. units for the parameters mentioned in the main text, this critical electric current is of the order of $10^{10}$ A/m$^2$, indicating a shallow trap. 
}
    \label{fig:trayectoria}
\end{figure}

Considering all the effects generated by the skyrmion–dislocation interaction, and taking into account the applied current, which plays a crucial role.
The current acts, on the one hand, on the dynamical part of the Lagrangian via spin-transfer torque, and, on the other hand, introduces a dissipative force associated with Rayleigh dissipation. These two contributions, expressed in stereographic coordinates, take the following form:
\begin{align}
   T=i\pi q(\dot{w}\bar{w}-\dot{\bar{w}}w)+iuq\pi(w+\bar{w})&& R=\pi \alpha  \dot{w}\dot{\bar{w}}-\pi\beta(\bar{u}\dot{w}+u\dot{\bar{w}})
\end{align}
Here,  $u = u_x + i u_y$. 
The collective dynamic description of the topological texture is summarized in Thiele´s equation\cite{Thiele1973, Troncoso2014, Troncoso2014b, Kim2023}, considering the presence of the Rayleigh dissipation:
\begin{equation}
    i\pi q\dot{w}=f(w-\xi,\bar{w}-\bar{\xi}) +iuq\pi+\pi u\beta-\pi\alpha \, \dot{w}
\end{equation}
where $\xi$ is the location (in complex coordinates) of the dislocation and $f$ contains the different effects generated by the interaction between spins and the DMI with the crystal dislocation, taking the following form:
$
f=-\partial_{\bar{w}}\,E_{EX}-\partial_{\bar{w}}\,E^{bulk}_{DMI}, 
$
where $E$ is the respective energy associated with the magnetic profile centered at $\zeta$.  Thus, solving this equation shows that, in the absence of dissipation and current ($\alpha=u=0$) the skyrmion undergoes circular motion around the dislocation (see the green dotted \ref{fig:trayectoria}.(a))
On the other hand, when dissipation is introduced into the system, the skyrmion, confined by the dislocation, precesses around it until reach the equilibrium position (see white or orange line).

To delocalize the skyrmion from its orbit, it is necessary to introduce an external perturbation, i.e, the electric current terms mentioned above.
In this way, it is possible to determine the critical current required to remove the skyrmion from the orbit produced by the DMI and exchange interaction potentials. To this end, the skyrmion's average velocity was calculated as a function of the applied current.

\section{Semiclassical theory of the skyrmion bound states in the dislocation-based traps}
Magnetic skyrmions, though often treated as classical topological spin textures, exhibit rich quantum mechanical behavior that becomes prominent at the nanoscale. Their collective dynamics can be quantized, giving rise to discrete magnon-skyrmion hybrid excitations and quantized modes analogous to Landau levels\cite{Girvin1999, Ezawa2013} for charged particles in a magnetic field. The nontrivial topology of the skyrmion induces an emergent electromagnetic field that couples to the spins, resulting in quantized transport phenomena, such as the topological Hall effect. Landau levels in magnetic skyrmions arise from the quantization of the cyclotron motion of the skyrmion's spin texture, which acts like an emergent magnetic field\cite{Weber2022, Li2020}. This creates discrete energy levels, similar to those in a traditional magnetic field, but with unique properties due to the skyrmion's topology. These emergent Landau levels are observed in the system's magnon spectra and can be used to understand and manipulate skyrmion behavior via techniques such as microwave absorption. Moreover, quantum fluctuations can stabilize or destabilize skyrmionic states, influence their tunneling between metastable configurations, and enable the formation of quantum superpositions of skyrmions and antiskyrmions\cite{RoldanMolina2015, Petrovi2025}. These effects position skyrmions as promising candidates for exploring quantum topological phenomena and as potential carriers of quantum information in spintronic systems.

\begin{figure}[H]
    \centering
    \includegraphics[width=1\linewidth]{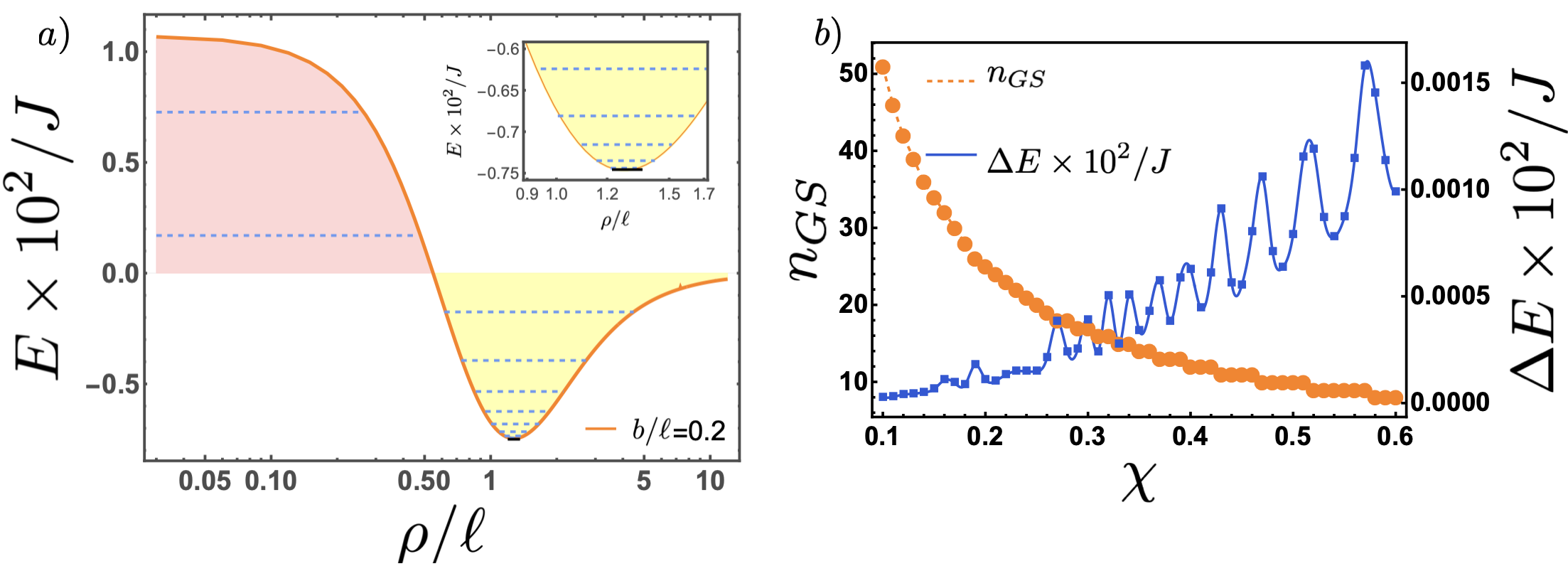}
    \caption{a) Energy $E\times 10^2/J$ as a function of $\rho/\ell$ for $b/\ell = 0.2$ and $\chi=0.5$. The dashed lines represent the confined eigenvalues $E_n$. The inset figure corresponds to a zoom of the curve around the minimum eigenstate $E_{10}$ (black line). b) $n_{GS}$ and $\Delta E= E_{n_{GS}}- E(\rho_{min})$ as a function of $\chi$.}
    \label{fig:EnergyQM}
\end{figure}

We start from the classical Lagrangian for the gyrotropic modes, which, in terms of the skyrmion number $q$,  acquires the form:
$$
L[w,\bar{w},\dot{w},\dot{\bar{w}},t]=iq\pi (\dot{w}\bar{w}-\dot{\bar{w}}w)-E(w,\bar{w}),
$$
the conjugated moments become: $p_w=iq\pi \bar{w}$ y $p_{\bar{w}}=-iq\pi w$. The canonical Poisson brackets are $\{w,p_w\}=1$, leading to $\{w,\bar{w}\}=-\frac{i}{q\pi}$. On the other hand the Hamiltonian becomes, simply, $H(w,\bar{w})=E(w,\bar{w})$. Upon quantization we obtain $[w,w^\dagger]=\frac{D}{q\pi J S}=\chi$
which leads to a noncommutative geometry, characterized by the algebra of a harmonic oscillator\cite{Ezawa2013}. Let us consider the states $|n\rangle$, such that $w^\dagger w|n\rangle=\chi n|n\rangle$. Since the physical properties are rotationally invariant within the plane, such basis is an eigenstate of the Hamiltonian $H(w,w^\dagger)=\frac{1}{2}(E(w,w^\dagger)+E(w^\dagger,w))$ with eigenvalues $E_{n} = \frac{1}{2}\left[E(\sqrt{\chi n}) + E(\sqrt{\chi (n + 1)})\right]$, plotted in Figure \ref{fig:EnergyQM}.
We note that these states are also eigenvalues of the $z$ component of the orbital angular momentum operator that, in complex coordinates, becomes $L_z=\pi q \bar{w}w$, which upon quantization becomes $L_n=\pi q\chi \left(n+\frac{1}{2}\right)$. The prefactor is nothing but $\hbar/(\varepsilon \tau)$, i.e. just Planck's constant in the reduced units used in this work. The emergence of half-integer orbital angular momentum is a curiosity, as they are forbidden in three-dimensional systems. Nevertheless, it has been reported and detected in other two-dimensional systems with topological terms in the action\cite{Kuleshov2016}. 
As is well known, the ground state wavefunction is represented\cite{Landau1981} by:
$$
\psi_{{GS}}(x)=\frac{1}{\sqrt{2^{n_{GS}}n_{GS}!\chi\sqrt{\pi}}}e^{-\frac{x^2}{2\chi}}H_{n_{GS}}(x)
$$
Here $H_n(x)$ represents the $n$-th Hermite polynomials\cite{Abramowitz1965, Olver2010}.

In the quantum theory, the skyrmion's coordinates fail to commute, which leads to an uncertainty relation $\Delta x\Delta y\gtrsim \chi$. This condition prevents us from a simple $xy$ representation of the skyrmion's wavefunction. Despite this, a representation is possible by using the Wigner transform $W(x,y)$\cite{Feynman1998, Wong1998}. The result for the Wigner function of the ground state is:
\begin{equation}
    W_n(x,y)=\frac{(-1)^n}{\pi\chi}e^{-\frac{x^2+y^2}{\chi}} \mathrm{L}_n\left(\frac{4\left(x^2+y^2\right)}{\chi}\right),
\end{equation} 
where $\mathrm{L}_n$ stands for the $n$-th  Laguerre polynomial\cite{Abramowitz1965, Olver2010}. These functions are presented in Appendix C.

Preparing skyrmions in well-defined energy eigenstates and quantum levels of well-defined orbital angular momentum is crucial for advancing their role as controllable quantum objects. In such states, skyrmions exhibit quantized dynamical properties that enable precise manipulation and robust coherence, key ingredients for quantum technologies. The quantization of their orbital angular momentum not only determines the symmetry and stability of the skyrmion textures but also governs their coupling to external fields, magnons, and conduction electrons. By isolating skyrmions in specific eigenstates, one can exploit their topological protection while accessing discrete, tunable degrees of freedom, opening pathways toward skyrmion-based qubits, quantum memories, and coherent spintronic devices that leverage the interplay between topology and quantum mechanics.

\section{Application: a dislocation-based skyrmion race track memory}
A fundamental concept derived from the behavior of a dislocation as a shallow trap is the development of a device that uses an arrangement of dislocations as trapping centers. In this configuration, a mobile skyrmion transitions between these centers, akin to the mechanism observed in race track memory systems. This concept is elaborated upon in recent studies, such as those in \cite{Yang2021, He2023}, which provide a detailed exploration of the dynamics and potential applications of such a system in advanced technological contexts. Owing to the shallowness of the trapping mechanism, the transition is regulated by an extremely small electric current that functions analogously to a guiding fret on a racetrack. The determination of the precise location of the skyrmion along the racetrack is achieved through the measurement of the topological Hall voltage induced by the skyrmion texture. This voltage arises from the presence of a skyrmion near its designated position along the track. This voltage is generated by the topological properties of the skyrmion, which interact with the material's electronic structure, thereby providing a reliable method for pinpointing the skyrmion's exact position within the racetrack system. 
The topological Hall effect (THE) arises as a direct manifestation of the real-space Berry curvature generated by skyrmion spin textures. When conduction electrons traverse a material hosting skyrmions, their spins adiabatically follow the local magnetization, acquiring a Berry phase equivalent to the flux of an emergent magnetic field produced by the nontrivial topology of the skyrmions\cite{Bruno2004, Qin2019, Zhang2022}. This emergent field acts as an additional Lorentz force, deflecting charge carriers and leading to a Hall voltage even in the absence of an external magnetic field or conventional spin-orbit coupling. The magnitude of the topological Hall signal thus encodes information about the density and topology of the underlying skyrmion lattice, serving as a key experimental signature of their presence\cite{Jiang2016}. This effect not only highlights the interplay between electronic transport and spin topology but also offers a pathway for detecting and manipulating skyrmions in spintronic applications.

\section{Summary and conclusions}
This article offers a comprehensive examination of the interplay between a screw dislocation in a perturbed magnetic lattice. The investigation focuses on the potential mechanisms of coupling and elucidates a significant connection between two distinct areas of materials science: topological magnetism and topological elasticity. This paper conducts a thorough analysis, aiming to integrate these fields by emphasizing the theoretical and experimental advances that facilitate a deeper understanding of the fundamental principles governing these phenomena.  The integration of these concepts significantly contributes to the scientific enterprise by providing fresh perspectives and opportunities for future investigations into the interactions and mutual influences between these two distinct yet interconnected disciplines. An in-depth examination of the fundamental principles governing dislocations and topological spin textures is conducted. This analysis extends to explore their influence on magnetic anisotropy and the Dzyaloshinskii-Moriya interaction, as well as the possible coupling between spin and lattice structures. Moreover, we derive a discrete Heisenberg exchange to leading order in a gradient expansion. The investigation provides a comprehensive understanding of these complex interconnections, which are pivotal in advancing the field of condensed matter physics. Subsequently, an in-depth analysis of the quantized dynamics of the skyrmion is conducted to identify its discrete quantum states. This study involves deriving an effective Thiele equation and using a Lagrangian formalism to understand and describe the behavior of these quantum entities comprehensively.  In conclusion, we aim to demonstrate the practical application of the concepts discussed in our paper through the design and implementation of devices that, while straightforward, remain highly effective. These devices utilize shallow traps formed by an array of dislocations, which function analogously to the grooves on a race track. This configuration effectively governs the movement of elements or particles via a low-current activation mechanism. By elaborating on these principles, we seek to provide a comprehensive understanding of how these ideas can be realized in tangible technological applications.

\section*{Acknowledgements}
Funding is acknowledged from Fondecyt Regular 1230515 and Cedenna CIA250002. M.A.C. acknowledges Proyecto ANID Fondecyt Postdoctorado 3240112.  S. Allende acknowledges DICYT regular 042431AP  .
\appendix
\section{Appendix A: Affine Metric}
In cartesian coordinates the metric is:
\begin{equation}
    g_{ij}= \begin{pmatrix}
1+\frac{b^2\sin^2(\theta)}{4 \pi^2 \rho^2} &
-\frac{b^2 \sin(2 \theta)}{8 \pi^2 \rho^2} &
-\frac{b \sin(\theta)}{2 \pi \rho} \\
-\frac{b^2 \sin(2 \theta)}{8 \pi^2 \rho^2} &
1+\frac{b^2\cos^2(\theta)}{8 \pi^2 \rho^2} &
\frac{b \cos(\theta)}{2 \pi \rho} \\
-\frac{b \sin(\theta)}{2 \pi \rho} &
\frac{b \cos(\theta)}{2 \pi \rho} &
1
\end{pmatrix}
\end{equation}
and the only non-vanishing components of the connection are:
\begin{equation}
\begin{aligned}
\Gamma^{z}_{xx} &= -\Gamma^{z}_{yy} = \frac{b\,\cos\theta\,\sin\theta}{\pi \rho^{2}}, \\
\Gamma^{z}_{xy} &= \Gamma^{z}_{yx} = -\frac{b\,\cos(2\theta)}{2\pi \rho^{2}}, 
%\Gamma^{z}_{yy} &= -\frac{b\,\cos\theta\,\sin\theta}{\pi r^{2}}.
\end{aligned}
\label{eq:GammaZ_polar}
\end{equation}

In complex coordinates, the inverse metric is:
\begin{equation}
    g^{ij}(\zeta,\bar{\zeta})=\begin{pmatrix}
1 &
0 &
-\frac{ib(\zeta-\bar{\zeta})}{4\pi \zeta\bar{\zeta}} \\
0 &
1 &
-\frac{b(\zeta+\bar{\zeta})}{4\pi \zeta\bar{\zeta}} \\
-\frac{ib(\zeta-\bar{\zeta})}{4\pi \zeta\bar{\zeta}} &
-\frac{b(\zeta+\bar{\zeta})}{4\pi \zeta\bar{\zeta}} &
1+\frac{b^2}{4\pi^2\zeta\bar{\zeta}}
\end{pmatrix}
\end{equation}

\section{Appendix B: Interfacial DMI}

Furthermore, for the interfacial case, we obtain:
\begin{equation}
    H^{int}_{DMI}=\frac{DS^2}{a}\int d^2x\sqrt{g}g^{ij}(\bz\times\be_i)\cdot(\bS\times \nabla_j \bS)
\end{equation}
while in complex coordinates it becomes:
\begin{align}
   H_{DMI}^{int}=\int id\zeta\wedge d\bar{\zeta}\left[\frac{2\partial_\zeta\psi}{(1+\psi\bar{\psi})^2}+\frac{bg^1\partial_\zeta\psi}{2\pi \zeta(1+\psi\bar{\psi})^2}-\frac{bg^1\bar{\psi}^2\partial_\zeta\psi}{2\pi\bar{\zeta}(1+\psi\bar{\psi})^2}+h.c.\right] 
\end{align}
\section{Appendix C: Wigner's function}
The results are presented in the following figure:
\begin{figure}
    \centering
    \includegraphics[width=1\linewidth]{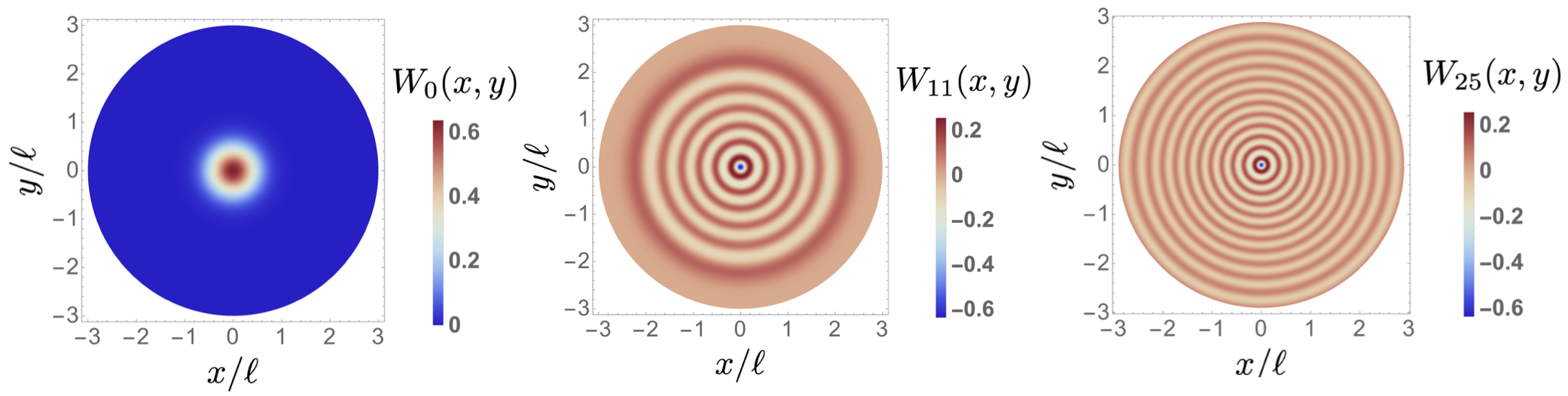}
    \caption{Density plots of the Wigner function for three energy levels ($n=0$, $n = 11$, and $n = 25$, ordered from left to right). The ground state corresponds to $n = 25$. We use $\chi=0.5$.}
    \label{fig:placeholder}
\end{figure}
\%bibliographystyle{unsrt} 
\bibliography{example}

\providecommand{\latin}[1]{#1}
\makeatletter
\providecommand{\doi}
  {\begingroup\let\do\@makeother\dospecials
  \catcode`\{=1 \catcode`\}=2 \doi@aux}
\providecommand{\doi@aux}[1]{\endgroup\texttt{#1}}
\makeatother
\providecommand*\mcitethebibliography{\thebibliography}
\csname @ifundefined\endcsname{endmcitethebibliography}  {\let\endmcitethebibliography\endthebibliography}{}
\begin{mcitethebibliography}{73}
\providecommand*\natexlab[1]{#1}
\providecommand*\mciteSetBstSublistMode[1]{}
\providecommand*\mciteSetBstMaxWidthForm[2]{}
\providecommand*\mciteBstWouldAddEndPuncttrue
  {\def\EndOfBibitem{\unskip.}}
\providecommand*\mciteBstWouldAddEndPunctfalse
  {\let\EndOfBibitem\relax}
\providecommand*\mciteSetBstMidEndSepPunct[3]{}
\providecommand*\mciteSetBstSublistLabelBeginEnd[3]{}
\providecommand*\EndOfBibitem{}
\mciteSetBstSublistMode{f}
\mciteSetBstMaxWidthForm{subitem}{(\alph{mcitesubitemcount})}
\mciteSetBstSublistLabelBeginEnd
  {\mcitemaxwidthsubitemform\space}
  {\relax}
  {\relax}

\bibitem[Dey and Roy(2021)Dey, and Roy]{Dey2021}
Dey,~P.; Roy,~J.~N. \emph{Spintronics}; Springer Singapore: Singapore, 2021; pp 75--101\relax
\mciteBstWouldAddEndPuncttrue
\mciteSetBstMidEndSepPunct{\mcitedefaultmidpunct}
{\mcitedefaultendpunct}{\mcitedefaultseppunct}\relax
\EndOfBibitem
\bibitem[Guo and Xi(2022)Guo, and Xi]{Guo2022}
Guo,~X.; Xi,~L. \emph{Spintronics}; Wiley, 2022; pp 221--242\relax
\mciteBstWouldAddEndPuncttrue
\mciteSetBstMidEndSepPunct{\mcitedefaultmidpunct}
{\mcitedefaultendpunct}{\mcitedefaultseppunct}\relax
\EndOfBibitem
\bibitem[Fert \latin{et~al.}(2017)Fert, Reyren, and Cros]{Fert2017}
Fert,~A.; Reyren,~N.; Cros,~V. Magnetic skyrmions: advances in physics and potential applications. \emph{Nature Reviews Materials} \textbf{2017}, \emph{2}\relax
\mciteBstWouldAddEndPuncttrue
\mciteSetBstMidEndSepPunct{\mcitedefaultmidpunct}
{\mcitedefaultendpunct}{\mcitedefaultseppunct}\relax
\EndOfBibitem
\bibitem[Li \latin{et~al.}(2023)Li, Wang, and Rasing]{Li2023}
Li,~S.; Wang,~X.; Rasing,~T. Magnetic skyrmions: Basic properties and potential applications. \emph{Interdisciplinary Materials} \textbf{2023}, \emph{2}, 260–289\relax
\mciteBstWouldAddEndPuncttrue
\mciteSetBstMidEndSepPunct{\mcitedefaultmidpunct}
{\mcitedefaultendpunct}{\mcitedefaultseppunct}\relax
\EndOfBibitem
\bibitem[Tokura and Kanazawa(2020)Tokura, and Kanazawa]{Tokura2020}
Tokura,~Y.; Kanazawa,~N. Magnetic Skyrmion Materials. \emph{Chemical Reviews} \textbf{2020}, \emph{121}, 2857–2897\relax
\mciteBstWouldAddEndPuncttrue
\mciteSetBstMidEndSepPunct{\mcitedefaultmidpunct}
{\mcitedefaultendpunct}{\mcitedefaultseppunct}\relax
\EndOfBibitem
\bibitem[Ji \latin{et~al.}(2024)Ji, Yang, Ahn, Moon, Ju, Im, Han, Lee, Park, Lee, Kim, and Hwang]{Ji2024}
Ji,~Y.; Yang,~S.; Ahn,~H.; Moon,~K.; Ju,~T.; Im,~M.; Han,~H.; Lee,~J.; Park,~S.; Lee,~C.; Kim,~K.; Hwang,~C. Direct Observation of Room‐Temperature Magnetic Skyrmion Motion Driven by Ultra‐Low Current Density in Van Der Waals Ferromagnets. \emph{Advanced Materials} \textbf{2024}, \emph{36}\relax
\mciteBstWouldAddEndPuncttrue
\mciteSetBstMidEndSepPunct{\mcitedefaultmidpunct}
{\mcitedefaultendpunct}{\mcitedefaultseppunct}\relax
\EndOfBibitem
\bibitem[Fullerton \latin{et~al.}(2025)Fullerton, Li, Solanki, Grebenchuk, Grzeszczyk, Chen, Šiškins, Novoselov, Koperski, Santos, and Phatak]{Fullerton2025}
Fullerton,~J.; Li,~Y.; Solanki,~H.; Grebenchuk,~S.; Grzeszczyk,~M.; Chen,~Z.; Šiškins,~M.; Novoselov,~K.~S.; Koperski,~M.; Santos,~E. J.~G.; Phatak,~C. Observation of Topological Chirality Switching Induced Freezing of a Skyrmion Crystal. \emph{Advanced Materials} \textbf{2025}, \relax
\mciteBstWouldAddEndPunctfalse
\mciteSetBstMidEndSepPunct{\mcitedefaultmidpunct}
{}{\mcitedefaultseppunct}\relax
\EndOfBibitem
\bibitem[Cacilhas \latin{et~al.}(2018)Cacilhas, Carvalho-Santos, Vojkovic, Carvalho, Pereira, Altbir, and N\'uñez]{Cacilhas2018}
Cacilhas,~R.; Carvalho-Santos,~V.~L.; Vojkovic,~S.; Carvalho,~E.~B.; Pereira,~A.~R.; Altbir,~D.; N\'uñez,~A.~S. Coupling of skyrmions mediated by the RKKY interaction. \emph{Applied Physics Letters} \textbf{2018}, \emph{113}\relax
\mciteBstWouldAddEndPuncttrue
\mciteSetBstMidEndSepPunct{\mcitedefaultmidpunct}
{\mcitedefaultendpunct}{\mcitedefaultseppunct}\relax
\EndOfBibitem
\bibitem[Koshibae \latin{et~al.}(2015)Koshibae, Kaneko, Iwasaki, Kawasaki, Tokura, and Nagaosa]{Koshibae2015}
Koshibae,~W.; Kaneko,~Y.; Iwasaki,~J.; Kawasaki,~M.; Tokura,~Y.; Nagaosa,~N. Memory functions of magnetic skyrmions. \emph{Japanese Journal of Applied Physics} \textbf{2015}, \emph{54}, 053001\relax
\mciteBstWouldAddEndPuncttrue
\mciteSetBstMidEndSepPunct{\mcitedefaultmidpunct}
{\mcitedefaultendpunct}{\mcitedefaultseppunct}\relax
\EndOfBibitem
\bibitem[Luo and You(2021)Luo, and You]{Luo2021}
Luo,~S.; You,~L. Skyrmion devices for memory and logic applications. \emph{APL Materials} \textbf{2021}, \emph{9}\relax
\mciteBstWouldAddEndPuncttrue
\mciteSetBstMidEndSepPunct{\mcitedefaultmidpunct}
{\mcitedefaultendpunct}{\mcitedefaultseppunct}\relax
\EndOfBibitem
\bibitem[Sisodia \latin{et~al.}(2022)Sisodia, Pelloux-Prayer, Buda-Prejbeanu, Anghel, Gaudin, and Boulle]{Sisodia2022}
Sisodia,~N.; Pelloux-Prayer,~J.; Buda-Prejbeanu,~L.~D.; Anghel,~L.; Gaudin,~G.; Boulle,~O. Robust and Programmable Logic-In-Memory Devices Exploiting Skyrmion Confinement and Channeling Using Local Energy Barriers. \emph{Physical Review Applied} \textbf{2022}, \emph{18}\relax
\mciteBstWouldAddEndPuncttrue
\mciteSetBstMidEndSepPunct{\mcitedefaultmidpunct}
{\mcitedefaultendpunct}{\mcitedefaultseppunct}\relax
\EndOfBibitem
\bibitem[Zhao \latin{et~al.}(2024)Zhao, Hua, Song, Yu, and Jiang]{Zhao2024}
Zhao,~L.; Hua,~C.; Song,~C.; Yu,~W.; Jiang,~W. Realization of skyrmion shift register. \emph{Science Bulletin} \textbf{2024}, \emph{69}, 2370–2378\relax
\mciteBstWouldAddEndPuncttrue
\mciteSetBstMidEndSepPunct{\mcitedefaultmidpunct}
{\mcitedefaultendpunct}{\mcitedefaultseppunct}\relax
\EndOfBibitem
\bibitem[Nakahara(2003)]{Nakahara2003}
Nakahara,~M. \emph{Geometry, Topology and Physics, Second Edition}; Graduate student series in physics; Taylor \& Francis, 2003\relax
\mciteBstWouldAddEndPuncttrue
\mciteSetBstMidEndSepPunct{\mcitedefaultmidpunct}
{\mcitedefaultendpunct}{\mcitedefaultseppunct}\relax
\EndOfBibitem
\bibitem[Zang \latin{et~al.}(2018)Zang, Cros, and Hoffmann]{Zang2018}
Zang,~J., Cros,~V., Hoffmann,~A., Eds. \emph{Topology in Magnetism}, 1st ed.; Springer Series in Solid-State Sciences; Springer International Publishing: Cham, Switzerland, 2018\relax
\mciteBstWouldAddEndPuncttrue
\mciteSetBstMidEndSepPunct{\mcitedefaultmidpunct}
{\mcitedefaultendpunct}{\mcitedefaultseppunct}\relax
\EndOfBibitem
\bibitem[Anderson \latin{et~al.}(2017)Anderson, Hirth, and Lothe]{Anderson2017}
Anderson,~P.~M.; Hirth,~J.~P.; Lothe,~J. \emph{Theory of Dislocations}, 3rd ed.; Cambridge University Press: Cambridge, England, 2017\relax
\mciteBstWouldAddEndPuncttrue
\mciteSetBstMidEndSepPunct{\mcitedefaultmidpunct}
{\mcitedefaultendpunct}{\mcitedefaultseppunct}\relax
\EndOfBibitem
\bibitem[Caillard(2007)]{Caillard2007}
Caillard,~D. Dislocations and Mechanical Properties. \emph{Alloy Physics} \textbf{2007}, 281–345\relax
\mciteBstWouldAddEndPuncttrue
\mciteSetBstMidEndSepPunct{\mcitedefaultmidpunct}
{\mcitedefaultendpunct}{\mcitedefaultseppunct}\relax
\EndOfBibitem
\bibitem[Friedel(2013)]{Friedel2013}
Friedel,~J. \emph{Dislocations: International series of monographs on solid state physics}; Elsevier: Amsterdam, Netherlands, 2013\relax
\mciteBstWouldAddEndPuncttrue
\mciteSetBstMidEndSepPunct{\mcitedefaultmidpunct}
{\mcitedefaultendpunct}{\mcitedefaultseppunct}\relax
\EndOfBibitem
\bibitem[Chen \latin{et~al.}(2017)Chen, Jian, Li, Chang, Ge, Hanus, Yang, Chen, Huang, Snyder, and Pei]{Chen2017}
Chen,~Z.; Jian,~Z.; Li,~W.; Chang,~Y.; Ge,~B.; Hanus,~R.; Yang,~J.; Chen,~Y.; Huang,~M.; Snyder,~G.~J.; Pei,~Y. Lattice Dislocations Enhancing Thermoelectric PbTe in Addition to Band Convergence. \emph{Advanced Materials} \textbf{2017}, \emph{29}\relax
\mciteBstWouldAddEndPuncttrue
\mciteSetBstMidEndSepPunct{\mcitedefaultmidpunct}
{\mcitedefaultendpunct}{\mcitedefaultseppunct}\relax
\EndOfBibitem
\bibitem[Han \latin{et~al.}(2025)Han, Dong, Yao, Zhang, Zhang, Gong, Huang, Gong, Wang, Zhang, Liu, Sun, Zhu, Li, Luo, Awaji, Wang, Xie, Hosono, and Ma]{Han2025}
Han,~M.; Dong,~C.; Yao,~C.; Zhang,~Z.; Zhang,~Q.; Gong,~Y.; Huang,~H.; Gong,~D.; Wang,~D.; Zhang,~X.; Liu,~F.; Sun,~Y.; Zhu,~Z.; Li,~J.; Luo,~J.; Awaji,~S.; Wang,~X.; Xie,~J.; Hosono,~H.; Ma,~Y. Asymmetric Stress Engineering of Dense Dislocations in Brittle Superconductors for Strong Vortex Pinning. \emph{Advanced Materials} \textbf{2025}, \relax
\mciteBstWouldAddEndPunctfalse
\mciteSetBstMidEndSepPunct{\mcitedefaultmidpunct}
{}{\mcitedefaultseppunct}\relax
\EndOfBibitem
\bibitem[Kaappa \latin{et~al.}(2024)Kaappa, Santa-aho, Honkanen, Vippola, and Laurson]{Kaappa2024}
Kaappa,~S.; Santa-aho,~S.; Honkanen,~M.; Vippola,~M.; Laurson,~L. Magnetic domain walls interacting with dislocations in micromagnetic simulations. \emph{Communications Materials} \textbf{2024}, \emph{5}\relax
\mciteBstWouldAddEndPuncttrue
\mciteSetBstMidEndSepPunct{\mcitedefaultmidpunct}
{\mcitedefaultendpunct}{\mcitedefaultseppunct}\relax
\EndOfBibitem
\bibitem[Kurtzig and Patel(1970)Kurtzig, and Patel]{Kurtzig1970}
Kurtzig,~A.; Patel,~J. Interaction of magnetic domain walls and individual dislocations. \emph{Physics Letters A} \textbf{1970}, \emph{33}, 123–125\relax
\mciteBstWouldAddEndPuncttrue
\mciteSetBstMidEndSepPunct{\mcitedefaultmidpunct}
{\mcitedefaultendpunct}{\mcitedefaultseppunct}\relax
\EndOfBibitem
\bibitem[Hu \latin{et~al.}(2018)Hu, Huang, Wang, Jiang, Ni, Zhou, Zielasek, Lagally, Huang, and Liu]{Hu2018}
Hu,~L.; Huang,~H.; Wang,~Z.; Jiang,~W.; Ni,~X.; Zhou,~Y.; Zielasek,~V.; Lagally,~M.; Huang,~B.; Liu,~F. Ubiquitous Spin-Orbit Coupling in a Screw Dislocation with High Spin Coherency. \emph{Physical Review Letters} \textbf{2018}, \emph{121}\relax
\mciteBstWouldAddEndPuncttrue
\mciteSetBstMidEndSepPunct{\mcitedefaultmidpunct}
{\mcitedefaultendpunct}{\mcitedefaultseppunct}\relax
\EndOfBibitem
\bibitem[Carpio \latin{et~al.}(2008)Carpio, Bonilla, Juan, and Vozmediano]{Carpio2008}
Carpio,~A.; Bonilla,~L.~L.; Juan,~F.~d.; Vozmediano,~M. A.~H. Dislocations in graphene. \emph{New Journal of Physics} \textbf{2008}, \emph{10}, 053021\relax
\mciteBstWouldAddEndPuncttrue
\mciteSetBstMidEndSepPunct{\mcitedefaultmidpunct}
{\mcitedefaultendpunct}{\mcitedefaultseppunct}\relax
\EndOfBibitem
\bibitem[Saji \latin{et~al.}(2025)Saji, Vidal-Silva, Nunez, and Troncoso]{Saji2025}
Saji,~C.; Vidal-Silva,~N.; Nunez,~A.~S.; Troncoso,~R.~E. Topological magnonic dislocation modes. \emph{Physical Review B} \textbf{2025}, \emph{111}\relax
\mciteBstWouldAddEndPuncttrue
\mciteSetBstMidEndSepPunct{\mcitedefaultmidpunct}
{\mcitedefaultendpunct}{\mcitedefaultseppunct}\relax
\EndOfBibitem
\bibitem[Yamada \latin{et~al.}(2022)Yamada, Li, Lin, Peterson, Hughes, and Bahl]{Yamada2022}
Yamada,~S.~S.; Li,~T.; Lin,~M.; Peterson,~C.~W.; Hughes,~T.~L.; Bahl,~G. Bound states at partial dislocation defects in multipole higher-order topological insulators. \emph{Nature Communications} \textbf{2022}, \emph{13}\relax
\mciteBstWouldAddEndPuncttrue
\mciteSetBstMidEndSepPunct{\mcitedefaultmidpunct}
{\mcitedefaultendpunct}{\mcitedefaultseppunct}\relax
\EndOfBibitem
\bibitem[Turski and Mińkowski(2009)Turski, and Mińkowski]{Turski2009}
Turski,~L.~A.; Mińkowski,~M. Spin wave interaction with topological defects. \emph{Journal of Physics: Condensed Matter} \textbf{2009}, \emph{21}, 376001\relax
\mciteBstWouldAddEndPuncttrue
\mciteSetBstMidEndSepPunct{\mcitedefaultmidpunct}
{\mcitedefaultendpunct}{\mcitedefaultseppunct}\relax
\EndOfBibitem
\bibitem[Gestrin and Sal’nikova(2012)Gestrin, and Sal’nikova]{Gestrin2012}
Gestrin,~S.~G.; Sal’nikova,~E.~A. Interaction of spin waves with dislocations in ferrodielectrics. \emph{Russian Physics Journal} \textbf{2012}, \emph{54}, 1177–1184\relax
\mciteBstWouldAddEndPuncttrue
\mciteSetBstMidEndSepPunct{\mcitedefaultmidpunct}
{\mcitedefaultendpunct}{\mcitedefaultseppunct}\relax
\EndOfBibitem
\bibitem[Azhar \latin{et~al.}(2022)Azhar, Kravchuk, and Garst]{Azhar2022}
Azhar,~M.; Kravchuk,~V.~P.; Garst,~M. Screw Dislocations in Chiral Magnets. \emph{Physical Review Letters} \textbf{2022}, \emph{128}\relax
\mciteBstWouldAddEndPuncttrue
\mciteSetBstMidEndSepPunct{\mcitedefaultmidpunct}
{\mcitedefaultendpunct}{\mcitedefaultseppunct}\relax
\EndOfBibitem
\bibitem[Petrović \latin{et~al.}(2025)Petrović, Psaroudaki, Fischer, Garst, and Panagopoulos]{Petrovi2025}
Petrović,~A.~P.; Psaroudaki,~C.; Fischer,~P.; Garst,~M.; Panagopoulos,~C. Colloquium : Quantum properties and functionalities of magnetic skyrmions. \emph{Reviews of Modern Physics} \textbf{2025}, \emph{97}\relax
\mciteBstWouldAddEndPuncttrue
\mciteSetBstMidEndSepPunct{\mcitedefaultmidpunct}
{\mcitedefaultendpunct}{\mcitedefaultseppunct}\relax
\EndOfBibitem
\bibitem[Roldán-Molina \latin{et~al.}(2015)Roldán-Molina, Santander, Nunez, and Fernández-Rossier]{RoldanMolina2015}
Roldán-Molina,~A.; Santander,~M.~J.; Nunez,~A.~S.; Fernández-Rossier,~J. Quantum fluctuations stabilize skyrmion textures. \emph{Physical Review B} \textbf{2015}, \emph{92}\relax
\mciteBstWouldAddEndPuncttrue
\mciteSetBstMidEndSepPunct{\mcitedefaultmidpunct}
{\mcitedefaultendpunct}{\mcitedefaultseppunct}\relax
\EndOfBibitem
\bibitem[Ornelas \latin{et~al.}(2025)Ornelas, Nape, de~Mello~Koch, and Forbes]{Ornelas2025}
Ornelas,~P.; Nape,~I.; de~Mello~Koch,~R.; Forbes,~A. Topological rejection of noise by quantum skyrmions. \emph{Nature Communications} \textbf{2025}, \emph{16}\relax
\mciteBstWouldAddEndPuncttrue
\mciteSetBstMidEndSepPunct{\mcitedefaultmidpunct}
{\mcitedefaultendpunct}{\mcitedefaultseppunct}\relax
\EndOfBibitem
\bibitem[Roldán-Molina \latin{et~al.}(2016)Roldán-Molina, Nunez, and Fernández-Rossier]{RoldanMolina2016}
Roldán-Molina,~A.; Nunez,~A.~S.; Fernández-Rossier,~J. Topological spin waves in the atomic-scale magnetic skyrmion crystal. \emph{New Journal of Physics} \textbf{2016}, \emph{18}, 045015\relax
\mciteBstWouldAddEndPuncttrue
\mciteSetBstMidEndSepPunct{\mcitedefaultmidpunct}
{\mcitedefaultendpunct}{\mcitedefaultseppunct}\relax
\EndOfBibitem
\bibitem[Zou \latin{et~al.}(2025)Zou, Bosco, Klinovaja, and Loss]{Zou2025}
Zou,~J.; Bosco,~S.; Klinovaja,~J.; Loss,~D. Topological spin textures enabling quantum transmission. \emph{Physical Review Research} \textbf{2025}, \emph{7}\relax
\mciteBstWouldAddEndPuncttrue
\mciteSetBstMidEndSepPunct{\mcitedefaultmidpunct}
{\mcitedefaultendpunct}{\mcitedefaultseppunct}\relax
\EndOfBibitem
\bibitem[Landau \latin{et~al.}(1984)Landau, Pitaevskii, Lifshitz, and Kosevich]{Landau1984}
Landau,~L.~D.; Pitaevskii,~L.~P.; Lifshitz,~E.~M.; Kosevich,~A.~M. \emph{Theory of elasticity}, 3rd ed.; Butterworth-Heinemann: Oxford, England, 1984\relax
\mciteBstWouldAddEndPuncttrue
\mciteSetBstMidEndSepPunct{\mcitedefaultmidpunct}
{\mcitedefaultendpunct}{\mcitedefaultseppunct}\relax
\EndOfBibitem
\bibitem[Kleinert(2007)]{Kleinert2007}
Kleinert,~H. \emph{Multivalued fields: In condensed matter, electromagnetism, and gravitation}; World Scientific Publishing: Singapore, Singapore, 2007\relax
\mciteBstWouldAddEndPuncttrue
\mciteSetBstMidEndSepPunct{\mcitedefaultmidpunct}
{\mcitedefaultendpunct}{\mcitedefaultseppunct}\relax
\EndOfBibitem
\bibitem[Kleinert(1989)]{kleinert1989}
Kleinert,~H. \emph{Gauge Fields in Condensed Matter: Stresses and defects}; Gauge Fields in Condensed Matter; World Scientific, 1989\relax
\mciteBstWouldAddEndPuncttrue
\mciteSetBstMidEndSepPunct{\mcitedefaultmidpunct}
{\mcitedefaultendpunct}{\mcitedefaultseppunct}\relax
\EndOfBibitem
\bibitem[Bausch \latin{et~al.}(1999)Bausch, Schmitz, and Turski]{Bausch1999}
Bausch,~R.; Schmitz,~R.; Turski,~{\L}.~A. Quantum motion of electrons in topologically distorted crystals. \emph{Annalen der Physik} \textbf{1999}, \emph{511}, 181--189\relax
\mciteBstWouldAddEndPuncttrue
\mciteSetBstMidEndSepPunct{\mcitedefaultmidpunct}
{\mcitedefaultendpunct}{\mcitedefaultseppunct}\relax
\EndOfBibitem
\bibitem[Zee(2013)]{Zee2013}
Zee,~A. \emph{Einstein gravity in a nutshell}; In a Nutshell; Princeton University Press: Princeton, NJ, 2013\relax
\mciteBstWouldAddEndPuncttrue
\mciteSetBstMidEndSepPunct{\mcitedefaultmidpunct}
{\mcitedefaultendpunct}{\mcitedefaultseppunct}\relax
\EndOfBibitem
\bibitem[Wald(1984)]{Wald1984}
Wald,~R.~M. \emph{General Relativity}; University of Chicago Press: Chicago, IL, 1984\relax
\mciteBstWouldAddEndPuncttrue
\mciteSetBstMidEndSepPunct{\mcitedefaultmidpunct}
{\mcitedefaultendpunct}{\mcitedefaultseppunct}\relax
\EndOfBibitem
\bibitem[Lin \latin{et~al.}(2014)Lin, Batista, and Saxena]{Lin2014}
Lin,~S.-Z.; Batista,~C.~D.; Saxena,~A. Internal modes of a skyrmion in the ferromagnetic state of chiral magnets. \emph{Physical Review B} \textbf{2014}, \emph{89}\relax
\mciteBstWouldAddEndPuncttrue
\mciteSetBstMidEndSepPunct{\mcitedefaultmidpunct}
{\mcitedefaultendpunct}{\mcitedefaultseppunct}\relax
\EndOfBibitem
\bibitem[Camley and Livesey(2023)Camley, and Livesey]{Camley2023}
Camley,~R.~E.; Livesey,~K.~L. Consequences of the Dzyaloshinskii-Moriya interaction. \emph{Surface Science Reports} \textbf{2023}, \emph{78}, 100605\relax
\mciteBstWouldAddEndPuncttrue
\mciteSetBstMidEndSepPunct{\mcitedefaultmidpunct}
{\mcitedefaultendpunct}{\mcitedefaultseppunct}\relax
\EndOfBibitem
\bibitem[Borisov \latin{et~al.}(2021)Borisov, Kvashnin, Ntallis, Thonig, Thunstr\"{o}m, Pereiro, Bergman, Sj\"{o}qvist, Delin, Nordstr\"{o}m, and Eriksson]{Borisov2021}
Borisov,~V.; Kvashnin,~Y.~O.; Ntallis,~N.; Thonig,~D.; Thunstr\"{o}m,~P.; Pereiro,~M.; Bergman,~A.; Sj\"{o}qvist,~E.; Delin,~A.; Nordstr\"{o}m,~L.; Eriksson,~O. Heisenberg and anisotropic exchange interactions in magnetic materials with correlated electronic structure and significant spin-orbit coupling. \emph{Physical Review B} \textbf{2021}, \emph{103}\relax
\mciteBstWouldAddEndPuncttrue
\mciteSetBstMidEndSepPunct{\mcitedefaultmidpunct}
{\mcitedefaultendpunct}{\mcitedefaultseppunct}\relax
\EndOfBibitem
\bibitem[Bazaliy \latin{et~al.}(1998)Bazaliy, Jones, and Zhang]{Bazaliy1998}
Bazaliy,~Y.~B.; Jones,~B.~A.; Zhang,~S.-C. Modification of the Landau-Lifshitz equation in the presence of a spin-polarized current in colossal- and giant-magnetoresistive materials. \emph{Physical Review B} \textbf{1998}, \emph{57}, R3213–R3216\relax
\mciteBstWouldAddEndPuncttrue
\mciteSetBstMidEndSepPunct{\mcitedefaultmidpunct}
{\mcitedefaultendpunct}{\mcitedefaultseppunct}\relax
\EndOfBibitem
\bibitem[Fernández-Rossier \latin{et~al.}(2004)Fernández-Rossier, Braun, Núñez, and MacDonald]{FernndezRossier2004}
Fernández-Rossier,~J.; Braun,~M.; Núñez,~A.~S.; MacDonald,~A.~H. Influence of a uniform current on collective magnetization dynamics in a ferromagnetic metal. \emph{Physical Review B} \textbf{2004}, \emph{69}\relax
\mciteBstWouldAddEndPuncttrue
\mciteSetBstMidEndSepPunct{\mcitedefaultmidpunct}
{\mcitedefaultendpunct}{\mcitedefaultseppunct}\relax
\EndOfBibitem
\bibitem[Li and Zhang(2003)Li, and Zhang]{Li2003}
Li,~Z.; Zhang,~S. Magnetization dynamics with a spin-transfer torque. \emph{Physical Review B} \textbf{2003}, \emph{68}\relax
\mciteBstWouldAddEndPuncttrue
\mciteSetBstMidEndSepPunct{\mcitedefaultmidpunct}
{\mcitedefaultendpunct}{\mcitedefaultseppunct}\relax
\EndOfBibitem
\bibitem[Tatara \latin{et~al.}(2008)Tatara, Kohno, and Shibata]{Tatara2008}
Tatara,~G.; Kohno,~H.; Shibata,~J. Microscopic approach to current-driven domain wall dynamics. \emph{Physics Reports} \textbf{2008}, \emph{468}, 213–301\relax
\mciteBstWouldAddEndPuncttrue
\mciteSetBstMidEndSepPunct{\mcitedefaultmidpunct}
{\mcitedefaultendpunct}{\mcitedefaultseppunct}\relax
\EndOfBibitem
\bibitem[Bausch \latin{et~al.}(1998)Bausch, Schmitz, and Turski]{Bausch1998}
Bausch,~R.; Schmitz,~R.; Turski,~{\L}.~A. Single-Particle Quantum States in a Crystal with Topological Defects. \emph{Physical Review Letters} \textbf{1998}, \emph{80}, 2257--2260\relax
\mciteBstWouldAddEndPuncttrue
\mciteSetBstMidEndSepPunct{\mcitedefaultmidpunct}
{\mcitedefaultendpunct}{\mcitedefaultseppunct}\relax
\EndOfBibitem
\bibitem[Niu \latin{et~al.}(2024)Niu, Kwon, Ma, Cheng, Ophus, Miao, Sun, Wu, Liu, Parkin, Won, Schmid, Ding, and Chen]{Niu2024}
Niu,~H.; Kwon,~H.~Y.; Ma,~T.; Cheng,~Z.; Ophus,~C.; Miao,~B.; Sun,~L.; Wu,~Y.; Liu,~K.; Parkin,~S. S.~P.; Won,~C.; Schmid,~A.~K.; Ding,~H.; Chen,~G. Reducing crystal symmetry to generate out-of-plane Dzyaloshinskii–Moriya interaction. \emph{Nature Communications} \textbf{2024}, \emph{15}\relax
\mciteBstWouldAddEndPuncttrue
\mciteSetBstMidEndSepPunct{\mcitedefaultmidpunct}
{\mcitedefaultendpunct}{\mcitedefaultseppunct}\relax
\EndOfBibitem
\bibitem[Robertson \latin{et~al.}(2020)Robertson, Agostino, Chen, Kang, Mascaraque, Garcia~Michel, Won, Wu, Schmid, and Liu]{Robertson2020}
Robertson,~M.; Agostino,~C.~J.; Chen,~G.; Kang,~S.~P.; Mascaraque,~A.; Garcia~Michel,~E.; Won,~C.; Wu,~Y.; Schmid,~A.~K.; Liu,~K. In-plane Néel wall chirality and orientation of interfacial Dzyaloshinskii-Moriya vector in magnetic films. \emph{Physical Review B} \textbf{2020}, \emph{102}\relax
\mciteBstWouldAddEndPuncttrue
\mciteSetBstMidEndSepPunct{\mcitedefaultmidpunct}
{\mcitedefaultendpunct}{\mcitedefaultseppunct}\relax
\EndOfBibitem
\bibitem[Garg \latin{et~al.}(2003)Garg, Kochetov, Park, and Stone]{Garg2003}
Garg,~A.; Kochetov,~E.; Park,~K.-S.; Stone,~M. Spin coherent-state path integrals and the instanton calculus. \emph{Journal of Mathematical Physics} \textbf{2003}, \emph{44}, 48–70\relax
\mciteBstWouldAddEndPuncttrue
\mciteSetBstMidEndSepPunct{\mcitedefaultmidpunct}
{\mcitedefaultendpunct}{\mcitedefaultseppunct}\relax
\EndOfBibitem
\bibitem[Timofeev and Aristov(2022)Timofeev, and Aristov]{Timofeev2022}
Timofeev,~V.~E.; Aristov,~D.~N. Magnon band structure of skyrmion crystals and stereographic projection approach. \emph{Physical Review B} \textbf{2022}, \emph{105}\relax
\mciteBstWouldAddEndPuncttrue
\mciteSetBstMidEndSepPunct{\mcitedefaultmidpunct}
{\mcitedefaultendpunct}{\mcitedefaultseppunct}\relax
\EndOfBibitem
\bibitem[Horley \latin{et~al.}(2009)Horley, Vieira, Sacramento, and Dugaev]{Horley2009}
Horley,~P.~P.; Vieira,~V.~R.; Sacramento,~P.~D.; Dugaev,~V.~K. Application of the stereographic projection to studies of magnetization dynamics described by the Landau–Lifshitz–Gilbert equation. \emph{Journal of Physics A: Mathematical and Theoretical} \textbf{2009}, \emph{42}, 315211\relax
\mciteBstWouldAddEndPuncttrue
\mciteSetBstMidEndSepPunct{\mcitedefaultmidpunct}
{\mcitedefaultendpunct}{\mcitedefaultseppunct}\relax
\EndOfBibitem
\bibitem[Thiele(1973)]{Thiele1973}
Thiele,~A.~A. Steady-State Motion of Magnetic Domains. \emph{Physical Review Letters} \textbf{1973}, \emph{30}, 230–233\relax
\mciteBstWouldAddEndPuncttrue
\mciteSetBstMidEndSepPunct{\mcitedefaultmidpunct}
{\mcitedefaultendpunct}{\mcitedefaultseppunct}\relax
\EndOfBibitem
\bibitem[Troncoso and N\'uñez(2014)Troncoso, and N\'uñez]{Troncoso2014}
Troncoso,~R.~E.; N\'uñez,~A.~S. Brownian motion of massive skyrmions in magnetic thin films. \emph{Annals of Physics} \textbf{2014}, \emph{351}, 850–856\relax
\mciteBstWouldAddEndPuncttrue
\mciteSetBstMidEndSepPunct{\mcitedefaultmidpunct}
{\mcitedefaultendpunct}{\mcitedefaultseppunct}\relax
\EndOfBibitem
\bibitem[Troncoso and Núñez(2014)Troncoso, and Núñez]{Troncoso2014b}
Troncoso,~R.~E.; Núñez,~A.~S. Thermally assisted current-driven skyrmion motion. \emph{Physical Review B} \textbf{2014}, \emph{89}\relax
\mciteBstWouldAddEndPuncttrue
\mciteSetBstMidEndSepPunct{\mcitedefaultmidpunct}
{\mcitedefaultendpunct}{\mcitedefaultseppunct}\relax
\EndOfBibitem
\bibitem[Kim(2023)]{Kim2023}
Kim,~B.~S. Generalizing Thiele equation. \emph{Journal of Physics: Condensed Matter} \textbf{2023}, \emph{35}, 425901\relax
\mciteBstWouldAddEndPuncttrue
\mciteSetBstMidEndSepPunct{\mcitedefaultmidpunct}
{\mcitedefaultendpunct}{\mcitedefaultseppunct}\relax
\EndOfBibitem
\bibitem[Girvin(1999)]{Girvin1999}
Girvin,~S.~M. The Quantum Hall Effect: Novel Excitations and Broken Symmetries. 1999; \url{https://arxiv.org/abs/cond-mat/9907002}\relax
\mciteBstWouldAddEndPuncttrue
\mciteSetBstMidEndSepPunct{\mcitedefaultmidpunct}
{\mcitedefaultendpunct}{\mcitedefaultseppunct}\relax
\EndOfBibitem
\bibitem[Ezawa(2013)]{Ezawa2013}
Ezawa,~Z.~F. \emph{Quantum hall effects: Recent theoretical and experimental developments (3rd edition)}, 3rd ed.; World Scientific Publishing: Singapore, Singapore, 2013\relax
\mciteBstWouldAddEndPuncttrue
\mciteSetBstMidEndSepPunct{\mcitedefaultmidpunct}
{\mcitedefaultendpunct}{\mcitedefaultseppunct}\relax
\EndOfBibitem
\bibitem[Weber \latin{et~al.}(2022)Weber, Fobes, Waizner, Steffens, Tucker, B\"{o}hm, Beddrich, Franz, Gabold, Bewley, Voneshen, Skoulatos, Georgii, Ehlers, Bauer, Pfleiderer, B\"{o}ni, Janoschek, and Garst]{Weber2022}
Weber,~T.; Fobes,~D.~M.; Waizner,~J.; Steffens,~P.; Tucker,~G.~S.; B\"{o}hm,~M.; Beddrich,~L.; Franz,~C.; Gabold,~H.; Bewley,~R.; Voneshen,~D.; Skoulatos,~M.; Georgii,~R.; Ehlers,~G.; Bauer,~A.; Pfleiderer,~C.; B\"{o}ni,~P.; Janoschek,~M.; Garst,~M. Topological magnon band structure of emergent Landau levels in a skyrmion lattice. \emph{Science} \textbf{2022}, \emph{375}, 1025–1030\relax
\mciteBstWouldAddEndPuncttrue
\mciteSetBstMidEndSepPunct{\mcitedefaultmidpunct}
{\mcitedefaultendpunct}{\mcitedefaultseppunct}\relax
\EndOfBibitem
\bibitem[Li and Kovalev(2020)Li, and Kovalev]{Li2020}
Li,~B.; Kovalev,~A.~A. Magnon Landau Levels and Spin Responses in Antiferromagnets. \emph{Physical Review Letters} \textbf{2020}, \emph{125}\relax
\mciteBstWouldAddEndPuncttrue
\mciteSetBstMidEndSepPunct{\mcitedefaultmidpunct}
{\mcitedefaultendpunct}{\mcitedefaultseppunct}\relax
\EndOfBibitem
\bibitem[Kuleshov \latin{et~al.}(2016)Kuleshov, Mur, Narozhny, and Lozovik]{Kuleshov2016}
Kuleshov,~V.~M.; Mur,~V.~D.; Narozhny,~N.~B.; Lozovik,~Y.~E. Topological Phase and Half-Integer Orbital Angular Momenta in Circular Quantum Dots. \emph{Few-Body Systems} \textbf{2016}, \emph{57}, 1103–1126\relax
\mciteBstWouldAddEndPuncttrue
\mciteSetBstMidEndSepPunct{\mcitedefaultmidpunct}
{\mcitedefaultendpunct}{\mcitedefaultseppunct}\relax
\EndOfBibitem
\bibitem[Landau and Lifshitz(1981)Landau, and Lifshitz]{Landau1981}
Landau,~L.~D.; Lifshitz,~E.~M. In \emph{Quantum mechanics}, 3rd ed.; Menzies,~J., Ed.; Butterworth-Heinemann: Oxford, England, 1981\relax
\mciteBstWouldAddEndPuncttrue
\mciteSetBstMidEndSepPunct{\mcitedefaultmidpunct}
{\mcitedefaultendpunct}{\mcitedefaultseppunct}\relax
\EndOfBibitem
\bibitem[Abramowitz and Stegun(1965)Abramowitz, and Stegun]{Abramowitz1965}
Abramowitz,~M., Stegun,~I.~A., Eds. \emph{Handbook of mathematical functions}; Dover Books on Mathematics; Dover Publications: Mineola, NY, 1965\relax
\mciteBstWouldAddEndPuncttrue
\mciteSetBstMidEndSepPunct{\mcitedefaultmidpunct}
{\mcitedefaultendpunct}{\mcitedefaultseppunct}\relax
\EndOfBibitem
\bibitem[Olver(2010)]{Olver2010}
Olver,~F. W.~J. \emph{{NIST} handbook of mathematical functions hardback and {CD-ROM}}; Cambridge University Press: Cambridge, England, 2010\relax
\mciteBstWouldAddEndPuncttrue
\mciteSetBstMidEndSepPunct{\mcitedefaultmidpunct}
{\mcitedefaultendpunct}{\mcitedefaultseppunct}\relax
\EndOfBibitem
\bibitem[Feynman(1998)]{Feynman1998}
Feynman,~R.~P. \emph{Statistical mechanics}; Frontiers in Physics; Westview Press: Philadelphia, PA, 1998\relax
\mciteBstWouldAddEndPuncttrue
\mciteSetBstMidEndSepPunct{\mcitedefaultmidpunct}
{\mcitedefaultendpunct}{\mcitedefaultseppunct}\relax
\EndOfBibitem
\bibitem[Wong(1998)]{Wong1998}
Wong,~M.~W. \emph{Weyl Transforms}, 1998th ed.; Universitext; Springer: New York, NY, 1998\relax
\mciteBstWouldAddEndPuncttrue
\mciteSetBstMidEndSepPunct{\mcitedefaultmidpunct}
{\mcitedefaultendpunct}{\mcitedefaultseppunct}\relax
\EndOfBibitem
\bibitem[Yang \latin{et~al.}(2021)Yang, Liang, Tseng, and Chen]{Yang2021}
Yang,~Y.-H.; Liang,~Y.-P.; Tseng,~C.-H.; Chen,~S.-H. Exploring Skyrmion Racetrack Memory for High Performance Full-Nonvolatile FTL. 2021 IEEE 10th Non-Volatile Memory Systems and Applications Symposium (NVMSA). 2021; p 1–6\relax
\mciteBstWouldAddEndPuncttrue
\mciteSetBstMidEndSepPunct{\mcitedefaultmidpunct}
{\mcitedefaultendpunct}{\mcitedefaultseppunct}\relax
\EndOfBibitem
\bibitem[He \latin{et~al.}(2023)He, Tomasello, Luo, Zhang, Nie, Carpentieri, Han, Finocchio, and Yu]{He2023}
He,~B.; Tomasello,~R.; Luo,~X.; Zhang,~R.; Nie,~Z.; Carpentieri,~M.; Han,~X.; Finocchio,~G.; Yu,~G. All-Electrical 9-Bit Skyrmion-Based Racetrack Memory Designed with Laser Irradiation. \emph{Nano Letters} \textbf{2023}, \emph{23}, 9482–9490\relax
\mciteBstWouldAddEndPuncttrue
\mciteSetBstMidEndSepPunct{\mcitedefaultmidpunct}
{\mcitedefaultendpunct}{\mcitedefaultseppunct}\relax
\EndOfBibitem
\bibitem[Bruno \latin{et~al.}(2004)Bruno, Dugaev, and Taillefumier]{Bruno2004}
Bruno,~P.; Dugaev,~V.~K.; Taillefumier,~M. Topological Hall Effect and Berry Phase in Magnetic Nanostructures. \emph{Physical Review Letters} \textbf{2004}, \emph{93}\relax
\mciteBstWouldAddEndPuncttrue
\mciteSetBstMidEndSepPunct{\mcitedefaultmidpunct}
{\mcitedefaultendpunct}{\mcitedefaultseppunct}\relax
\EndOfBibitem
\bibitem[Qin \latin{et~al.}(2019)Qin, Liu, Lin, Shu, Xie, Lim, Li, He, Chow, and Chen]{Qin2019}
Qin,~Q.; Liu,~L.; Lin,~W.; Shu,~X.; Xie,~Q.; Lim,~Z.; Li,~C.; He,~S.; Chow,~G.~M.; Chen,~J. Emergence of Topological Hall Effect in a SrRuO3 Single Layer. \emph{Advanced Materials} \textbf{2019}, \emph{31}\relax
\mciteBstWouldAddEndPuncttrue
\mciteSetBstMidEndSepPunct{\mcitedefaultmidpunct}
{\mcitedefaultendpunct}{\mcitedefaultseppunct}\relax
\EndOfBibitem
\bibitem[Zhang \latin{et~al.}(2022)Zhang, Liu, Zhang, Yuan, Wen, Li, Zheng, Zhang, Hou, Yin, Liu, Peng, and Zhang]{Zhang2022}
Zhang,~C.; Liu,~C.; Zhang,~J.; Yuan,~Y.; Wen,~Y.; Li,~Y.; Zheng,~D.; Zhang,~Q.; Hou,~Z.; Yin,~G.; Liu,~K.; Peng,~Y.; Zhang,~X. Room‐Temperature Magnetic Skyrmions and Large Topological Hall Effect in Chromium Telluride Engineered by Self‐Intercalation. \emph{Advanced Materials} \textbf{2022}, \emph{35}\relax
\mciteBstWouldAddEndPuncttrue
\mciteSetBstMidEndSepPunct{\mcitedefaultmidpunct}
{\mcitedefaultendpunct}{\mcitedefaultseppunct}\relax
\EndOfBibitem
\bibitem[Jiang \latin{et~al.}(2016)Jiang, Zhang, Yu, Zhang, Wang, Benjamin Jungfleisch, Pearson, Cheng, Heinonen, Wang, Zhou, Hoffmann, and te Velthuis]{Jiang2016}
Jiang,~W.; Zhang,~X.; Yu,~G.; Zhang,~W.; Wang,~X.; Benjamin Jungfleisch,~M.; Pearson,~J.; Cheng,~X.; Heinonen,~O.; Wang,~K.~L.; Zhou,~Y.; Hoffmann,~A.; te Velthuis,~S. Direct observation of the skyrmion Hall effect. \emph{Nature Physics} \textbf{2016}, \emph{13}, 162–169\relax
\mciteBstWouldAddEndPuncttrue
\mciteSetBstMidEndSepPunct{\mcitedefaultmidpunct}
{\mcitedefaultendpunct}{\mcitedefaultseppunct}\relax
\EndOfBibitem
\end{mcitethebibliography}



%% For example:

%% \bibitem[Aladro et al.(2015)]{Aladro15} Aladro, R., Martín, S., Riquelme, D., et al. 2015, \aas, 579, A101

%%\end{thebibliography}

\end{document}